\documentclass{emulateapj}

\begin{document}

\title{Reexamination of the radial abundance Gradient Break in NGC 3359}
\author{H. J. Zahid and F. Bresolin}
\affil{Institute for Astronomy 2680 Woodlawn Dr., Honolulu, HI 96822}

%\begin{document}
\begin{abstract}
In this contribution, we reexamine the radial oxygen abundance gradient in the strongly barred spiral galaxy NGC~3359, for which, using an imaging spectrophotometric technique, Martin \& Roy detected a break  near the effective radius of the galaxy. We have new emission line flux measurements of HII regions in NGC 3359 from spectra obtained with the Subaru telescope to further investigate this claim. We find that there are small systematic variations in the line ratios determined from narrow-band imaging as compared to our spectroscopic measurements. We derive and apply a correction to the line ratios found by Martin \& Roy and statistically examine the validity of the gradient break proposed for NGC 3359 using recently developed metallicity diagnostics. We find that, with a high degree of confidence, a model with a  break fits the data significantly better than one without it. This suggests that the presence of a strong bar in spiral galaxies can generate measurable changes in the radial distribution of metals.
\end{abstract}

\section{Introduction}

Heavy metals are formed in stars and dispersed into the interstellar medium via mass loss processes leading to chemical enrichment. Spectral analysis of HII regions within galaxies can provide a measure of the chemical abundance \citep{Pagel1979}. The chemical abundance reflects the star formation history and the radial distribution of the chemical abundances in galaxies provides a key diagnostic for understanding galaxy formation and evolution \citep{Vila-Costas1992, Zaritsky1994}. A negative O/H gradient has been established in a large number of normal and barred spiral galaxies \citep[among others]{Searle1971, Garnett1987, Bresolin2009}. This is thought to be a result of greater gas processing in the center of these galaxies.

It has been found that the O/H abundance in barred galaxies is considerably more shallow than the gradient of normal spiral galaxies and this flattening is correlated to the strength of the bar \citep{Martin1994}. This has been attributed to homogenization of the abundances by large scale gas flows \citep{Pagel1979, Alloin1981}. Simulations by \citet{Friedli1994} have demonstrated that within the corotation radius gas accretion leads to increased star formation, resulting in an increase in the chemical abundance, while large-scale gas motions leads to a mixing outside the corotation radius. This mixing homogenizes the outer part of the galaxy resulting in a break at the corotation radius in the radial abundance gradient. This break is therefore taken as an indicator of a recently formed bar \citep{Roy1997, Friedli1999, Considere2000}.

NGC 3359 is a barred spiral galaxy classified as a SBc(rs)II by \citet{deVaucouleurs1991}. The bar ratio has been determined to be b/a = 0.3, making NGC 3359 a very strongly barred galaxy \citep{Martin1995b}. \citet[herafter MR95]{Martin1995} have determined the abundance gradient for NGC 3359 using the $[OIII]/H\beta$ and $[NII]/[OIII]$ abundance indicators obtained from monochromatic images of nebular emission lines of 77 HII regions. MR95 measure a steep slope, determined from a linear regression, for the O/H gradient inside the corotation radius of 4.7 kpc ($\Delta log(O/H)/\Delta R = -0.070 \pm 0.010 $ dex kpc$^{-1}$) and a flattening outside ($\Delta log(O/H)/\Delta R = -0.006 \pm 0.018 $ dex kpc$^{-1}$) which they attribute to the homogenization of the abundance due to gas mixing. To date there exists only a few examples in the literature of such breaks and this break in the slope observed by MR95 in NGC 3359 is one the clearest examples. Though models suggest that the formation and presence of bars result in a break in the abundance gradient, the cause and validity of the breaks are still not well established observationally. 

It has been argued by \citet{Pilyugin2003} that these bends may be attributable to systematic errors or improper use of the relationship between abundances and strong line intensities. He asserts that abundance indicators constructed from the oxygen strong-lines reproduces the metallicities for the high-abundance HII regions, but overestimates the metallicity in the low-abundance peripheral regions of galaxies leading to a false bend in the gradient. Moreover, he finds that a false bend in the abundance gradient may result from a bend in the ionization parameter gradient. 

We reexamine the abundance gradient of NGC 3359 and apply recently developed diagnostics appropriately calibrated and suited for this study. In $\S2$ we present our data and methods, in $\S3$ we compare our spectroscopic line ratio measurements to the photometric measurements of MR95. In $\S4$ we compare a model with two lines and a break with a single line model by performing a likelihood ratio test. In $\S5$ we examine the ionization parameter and in $\S6$ we present a brief discussion and our conclusions. In this work, we use a galactic distance of 14.4 Mpc for NGC 3359 determined from the radial velocity of 1008 km s$^{-1}$, using H$_0$ = 70 km s$^{-1}$ Mpc$^{-1}$. Folowing \citet{Ball1986}, we adopt $51^{\circ}$ and $172^{\circ}$ for the inclination and position angle of NGC 3359, respectively.

\section{Data}

\begin{figure}
\includegraphics[width=\columnwidth]{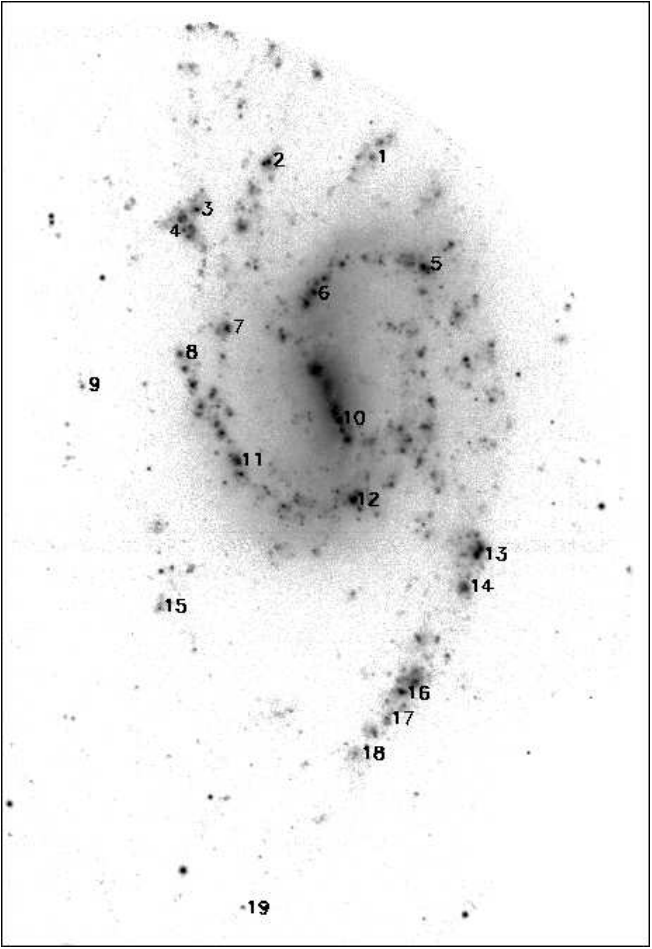}
\caption{H$\alpha$ image of NGC 3359 taken with FOCAS on-board the 8.2m Subaru telescope. A prominent bar,  showing a chain of HII regions, dominates the $H\alpha$ emission. Two distinct spiral arms can be seen with numerous HII regions highlighting the spiral structure. The 19 regions that have been observed spectroscopically and that meet our S/N cut are identified in the image. The image is oriented with North at the top and East to the left.}
\label{fig:img}
\end{figure}

\begin{deluxetable}{lccccc}
%\begin{deluxetable*}{lcccccccccccc}

\tablewidth{\columnwidth}

\tablecaption{HII Region Properties}

\tablehead{\colhead{ID} & \colhead{$R/R_{\mbox{\em\scriptsize eff}}$} & $R/R_{25}$ & \colhead{$R$ (kpc)} &\colhead{$H\beta/H\gamma$} & \colhead{E(B$-$V)} }

\startdata

1&1.17&0.47&7.18&2.70&0.52\\
2&1.19&0.48&7.31&2.43&0.28\\
3&1.35&0.54&8.29&2.50&0.35\\
4&1.36&0.55&8.38&2.41&0.27\\
5&0.86&0.34&5.26&2.58&0.42\\
6&0.47&0.19&2.91&2.77&0.58\\
7&0.82&0.33&5.05&2.75&0.56\\
8&1.11&0.45&6.85&2.62&0.45\\
9&1.80&0.72&11.0&2.82&0.62\\
10&0.16&0.06&0.95&2.47&0.33\\
11&0.69&0.28&4.25&2.43&0.29\\
12&0.57&0.23&3.49&2.41&0.27\\
13&1.42&0.57&8.73&2.48&0.33\\
14&1.45&0.58&8.92&2.33&0.19\\
15&1.49&0.60&9.15&2.12&0.00\\
16&1.57&0.63&9.63&2.45&0.30\\
17&1.64&0.66&10.0&2.20&0.07\\
18&1.70&0.68&10.4&\nodata&0.33\\
19&2.39&0.96&14.7&\nodata&0.33\\

\enddata

\label{tab:prop}

\tablecomments{Properties of the 19 HII regions from spectroscopic line flux measurements for NGC 3359. The first column gives the ID corresponding to the HII regions labeled in Figure \ref{fig:img}. Column 2 and 3 give the deprojected radial distances relative to the effective ($R_{\mbox{\em\scriptsize eff}} = 1.4$ arcmin) and 25th $B$ magnitude isophotal ($R_{25} = 3.6$ arcmin) radii, respectively, taken from \citet{deVaucouleurs1991}. Column 4 gives the deprojected galactocentric distances in kpc assuming a distance of 14.4 Mpc. Column 5 and 6 give the Balmer decrement determined from the $H\beta/H\gamma$ ratio and and extinction inferred from the reddening law of \citet{Cardelli1989}.}

\end{deluxetable}

%%%%%%%%%%%%%%%%%%%%%%%%%%%%%%%%%%%%%%%%%%%%%%%%%%%%%%%%%%%%%%%%%%%

\begin{deluxetable*}{lrrrrrrrcc}
%\begin{deluxetable*}{lcccccccccccc}

\tablewidth{475pt}

\tablecaption{HII Region Emission Line Strengths}

\tablehead{\colhead{ID} & \colhead{[OII]} &  \colhead{[OIII]}& \colhead{[OIII]}& \colhead{[OIII]}& \colhead{[NII]} & \colhead{[SII]}& \colhead{[SII]} & \colhead{12+log(O/H)}& \colhead{12+log(O/H)}  \\
&\colhead{3727}&\colhead{4363}&\colhead{4959}&\colhead{5007}&\colhead{6583}&\colhead{6717}&\colhead{6731}&(N2)&(O3N2)}

\startdata

1& 582 $\pm$ 36\,&\nodata& 60 $\pm$ 4.2& 184 $\pm$ 24\,& 31 $\pm$ 2.7& 36 $\pm$ 2.4& 25 $\pm$ 2.3&8.30$\, \pm \,$0.02&8.34$\, \pm \,$0.02\\
2& 245 $\pm$ 29\,& 1.7 $\pm$ 0.3& 130 $\pm$ 6.8& 389 $\pm$ 20\,& 18 $\pm$ 1.1& 20 $\pm$ 1.1& 16 $\pm$ 1.7&8.20$\, \pm \,$0.01&8.16$\, \pm \,$0.01\\
3& 205 $\pm$ 31\,& 2.8 $\pm$ 0.6& 138 $\pm$ 7.4& 406 $\pm$ 20\,& 10 $\pm$ 0.7& 9.3 $\pm$ 0.5& 6.8 $\pm$ 0.5&8.11$\, \pm \,$0.02&8.08$\, \pm \,$0.01\\
4&\nodata&\nodata& 44 $\pm$ 2.9& 135 $\pm$ 7.4& 36 $\pm$ 2.6& 38 $\pm$ 2.7& 27 $\pm$ 2.1&8.33$\, \pm \,$0.02&8.40$\, \pm \,$0.01\\
5& 231 $\pm$ 18\,&\nodata& 58 $\pm$ 3.1& 173 $\pm$ 9.2& 43 $\pm$ 2.3& 32 $\pm$ 1.7& 24 $\pm$ 1.5&8.38$\, \pm \,$0.02&8.39$\, \pm \,$0.01\\
6& 314 $\pm$ 63\,&\nodata& 13 $\pm$ 1.7& 41 $\pm$ 3.4& 53 $\pm$ 3.0& 38 $\pm$ 2.3& 27 $\pm$ 1.6&8.44$\, \pm \,$0.02&8.62$\, \pm \,$0.01\\
7& 287 $\pm$ 23\,&\nodata& 71 $\pm$ 3.7& 207 $\pm$ 10\,& 26 $\pm$ 1.4& 22 $\pm$ 1.2& 16 $\pm$ 0.8&8.27$\, \pm \,$0.01&8.30$\, \pm \,$0.01\\
8& 490 $\pm$ 64\,&\nodata& 29 $\pm$ 2.2& 74 $\pm$ 4.4& 36 $\pm$ 2.4& 32 $\pm$ 2.3& 23 $\pm$ 1.7&8.33$\, \pm \,$0.02&8.49$\, \pm \,$0.01\\
9& 410 $\pm$ 59\,&\nodata& 78 $\pm$ 6.8& 223 $\pm$ 17\,& 22 $\pm$ 4.1& 19 $\pm$ 4.8& 18 $\pm$ 4.6&8.23$\, \pm \,$0.04&8.26$\, \pm \,$0.03\\
10& 116 $\pm$ 11\,& & 17 $\pm$ 2.5& 60 $\pm$ 4.0& 71 $\pm$ 3.9& 38 $\pm$ 2.0& 28 $\pm$ 1.6&8.54$\, \pm \,$0.03&8.61$\, \pm \,$0.01\\
11& \nodata&\nodata& 51 $\pm$ 2.6& 152 $\pm$ 7.7& 39 $\pm$ 2.2& 35 $\pm$ 1.8& 25 $\pm$ 1.3&8.35$\, \pm \,$0.02&8.40$\, \pm \,$0.01\\
12& 272 $\pm$ 15\,&\nodata& 45 $\pm$ 2.4& 135 $\pm$ 7.0& 40 $\pm$ 2.2& 27 $\pm$ 1.5& 19 $\pm$ 1.0&8.36$\, \pm \,$0.02&8.42$\, \pm \,$0.01\\
13& 270 $\pm$ 17\,& 2.7 $\pm$ 0.4& 91 $\pm$ 4.8& 270 $\pm$ 13\,& 20 $\pm$ 1.1& 27 $\pm$ 1.4& 19 $\pm$ 1.4&8.22$\, \pm \,$0.01&8.23$\, \pm \,$0.01\\
14& 382 $\pm$ 31\,&\nodata& 64 $\pm$ 3.6& 191 $\pm$ 10\,& 32 $\pm$ 2.0& 40 $\pm$ 2.3& 28 $\pm$ 1.7&8.31$\, \pm \,$0.02&8.34$\, \pm \,$0.01\\
15&\nodata&\nodata& 111 $\pm$ 6.3& 331 $\pm$ 17\,& 23 $\pm$ 7.1& 38 $\pm$ 6.6& 33 $\pm$ 8.6&8.24$\, \pm \,$0.24&8.22$\, \pm \,$0.05\\
16&\nodata&\nodata & 85 $\pm$ 4.5& 250 $\pm$ 12\,& 28 $\pm$ 1.6& 41 $\pm$ 2.2& 31 $\pm$ 1.7&8.28$\, \pm \,$0.02&8.28$\, \pm \,$0.01\\
17& 332 $\pm$ 67\,& \nodata& 48 $\pm$ 4.5& 156 $\pm$ 9.3& 29 $\pm$ 3.9& 46 $\pm$ 4.5& 40 $\pm$ 4.4&8.29$\, \pm \,$0.03&8.35$\, \pm \,$0.02\\
18&\nodata&\nodata& 62 $\pm$ 11\,& 203 $\pm$ 20\,& 38 $\pm$ 12\,& 96 $\pm$ 30\,& 79 $\pm$ 36\,&8.35$\, \pm \,$0.18&8.35$\, \pm \,$0.11\\
19&\nodata&\nodata& 102 $\pm$ 12\,& 287 $\pm$ 22\,& 17 $\pm$ 5.9& 21 $\pm$ 5.3& 13 $\pm$ 2.6&8.19$\, \pm \,$0.39&8.19$\, \pm \,$0.11
\enddata

\label{tab:lr}

\tablecomments{Emission line strengths and metallicities of the 19 HII regions from spectroscopic line flux measurements for NGC~3359. Column 1 gives the ID corresponding to the HII regions labeled in Figure \ref{fig:img}. Columns 2-8 give the line flux relative to $H\beta$ of 100. Column 9 and 10 give the oxygen abundance, 12+log(O/H), derived from the $N2$ and $O3N2$ line ratios calibrated by \citet{Pettini2004}. The errors in the oxygen abundance are determined from propagating the uncertainties in the line ratios using a Monte Carlo method and do not reflect the systematic uncertainties of the strong line methods for abundance determinations. \citet{Kewley2008} provide a more detailed discussion of the systematic uncertainties.}

\end{deluxetable*}

NGC 3359 was observed using the Faint Object Camera and Spectrograph (FOCAS) on-board the 8.2m Subaru telescope on Mauna Kea \citep{Kashikawa2002}. Multi-object spectroscopy of 26 HII regions ranging from the center out to the periphery of the galaxy was taken. We obtained the blue portion of the spectrum from two 30-minute exposures using the 300R grism in second order, and the red portion of the spectrum from two 20-minute exposures using the VPH600 grism. The observations were made under photometric conditions with a seeing of $\sim\!1.7"$. The nominal spectral coverage for the blue portion is typically $\rm 3700 - 5100~ \AA$, although in a few spectra the $[OII]$ doublet centered at $\rm 3727~\AA$ is either not observed due to restricted wavelength coverage or lost in the noise. The nominal spectral coverage for the red portion of the spectra is $\rm 6300 - 8200~\AA$. The spectral resolution of these spectra is $\rm \sim5~\AA$.

Figure \ref{fig:img} shows an $H\alpha$ image of NGC 3359 obtained as part of the preimaging for this project. The intense star formation in NGC 3359 is traced out by its $\sim100$ HII regions. There is a prominent bar with a misaligned chain of HII regions along with two spiral arms. 

The spectroscopic data were reduced and 1D spectra were obtained using a suite of IDL routines developed specifically for these data. For each region, we perform an individual sky level subtraction by averaging the off-source pixels in each slit. The data were flux calibrated in IRAF\footnote{IRAF is distributed by the National Optical Astronomy Observatories, which are operated by the Association of Universities for Research in Astronomy, Inc., under cooperative agreement with the National Science Foundation.} using spectra of four observed standard stars. The fluxes for the emission lines were determined by summing all the pixels in the line bandpass while subtracting out the continuum determined from a linear fit to a region immediately outside the bandpass. 

The errors in the emission line flux, $\sigma_{l}$, were determined using the equation given in \citet{Gonzalez1994}:
\begin{equation}
\sigma_{l} = \sigma_{c}N^{1/2}[1 + EW/(N\Delta)]^{1/2}.
\label{eq:err}
\end{equation}
The error in the continuum, $\sigma_c$, is measured in the region near the emission line over which the linear fit to the continuum was performed. $N$ is the width of the region in pixels over which the flux in the emission line is measured. EW is the equivalent width of the emission line and $\Delta$ is the spectral resolution in \AA\ pixel$^{-1}$. The first term in the equation accounts for the uncertainty in the continuum near each line by measuring the rms of the continuum and multiplying by the width of the `base' of each emission line. The second term accounts for the Poisson error in the line intensity, which is expressed here through the equivalent width and is added in quadrature. For this work, where the S/N in our lines of interest is quite high, we consider this adequate for characterizing the line flux uncertainties. We account for additional uncertainties from the flat fielding, flux calibration and scaling due to the spectra being obtained through different grisms by adding in quadrature relative uncertainties of $1\%$, $4\%$ and $3\%$, respectively.

The data were dereddened using the reddening law of \citet{Cardelli1989} with the E(B$-$V) value determined from the Balmer decrement using the $H\beta/H\gamma$ ratio. For Case B recombination (T$_e$ = 10$^4$ K, $n_e$ = 100 cm$^{-3}$), this ratio is equal to 2.14 \citep{Hummer1987}. For regions 18 and 19 we were not able determine the Balmer decrement due to non-detection of the $H\gamma$ line. For these regions we have adopted the median value of the extinction. After dereddening the line fluxes, the spectra obtained in the two grisms were scaled to have the $H\alpha/H\beta$ ratio consistent with Case B recombination. We select our final sample of 19 HII regions by imposing a minimum signal-to-noise (S/N) ratio cut of 3 for the line ratios of interest. 

\citet{Pettini2004} derive an empirical method for estimating oxygen abundances based on the $[NII]/H\alpha$ and $([OIII]/H\beta)/([NII]/H\alpha)$ line ratios. They parameterize the relationship between $[NII]/H\alpha$ and $12 + log(O/H)$ with a third degree polynomial given as
\begin{equation}
12 + log(O/H) = 9.37 + 2.03 \cdot N2 + 1.26 \cdot N2^2 + 0.32 \cdot N2^3,
\end{equation}
where $N2 = log([NII]/H\alpha)$. Additionally, they derive a linear relationship between $([OIII]/H\beta)/([NII]/H\alpha)$ and $12 + log(O/H)$ given by
\begin{equation}
12 + log(O/H) = 8.73 - 0.32 \cdot O3N2, 
\end{equation}
where $O3N2 = log\{([OIII]/H\beta)/([NII]/H\alpha)\}$. These diagnostics are calibrated using HII regions where metallicity has been determined from either direct measurements of the electron temperature or by detailed photoionization modeling. The diagnostics using the $N2$ and $O3N2$ line ratios are monotonically increasing with O/H and are not double valued like many of the diagnostics using strong oxygen lines (i.e.~$R23$). Additionally, these diagnostics are calibrated over the full metallicity range observed in the HII regions of NGC 3359. The $1\sigma$ dispersion of the $N2$ and $O3N2$ calibrations are 0.18 and 0.14 dex, respectively.

Table \ref{tab:prop} lists some of the properties of the HII regions identified in Figure \ref{fig:img} and Table \ref{tab:lr} gives the emission line strengths and metallicities. The error estimates for metallicities given in Table 2 only reflect the statistical uncertainties, the systematic uncertainties in the absolute metallicities are considerably larger. However, we note that for the metallicity analysis performed in this study, we only require relative accuracy which the diagnostics do provide \citep{Kewley2008}.

\section{Spectroscopic vs. Photometric Line Ratios}

\begin{figure*}
\includegraphics[width=7in]{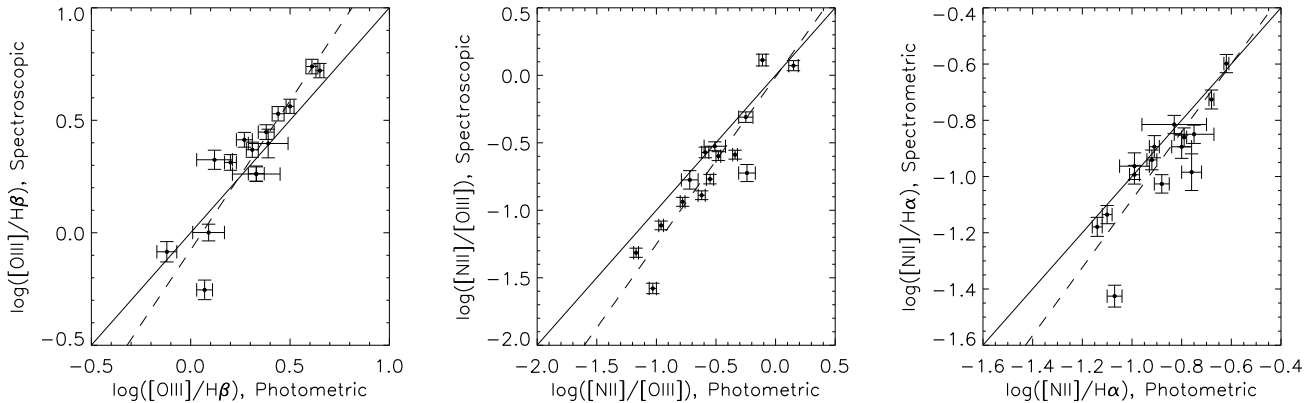}
\caption{The photometric measurements of MR95 compared with our spectroscopic measurements. The figure shows the $[OIII]/H\beta$ (left), $[NII]/[OIII]$ (center) and $[NII]/H\alpha$ (right) line ratios. The data points are the 17 regions for which there is both a photometric and spectroscopic line ratio measurement. The solid line in each of the plots is the one-to-one agreement and the dashed line is a least-square fit to the data. The error bars are the statistical uncertainties taken from MR95 (x-direction) and this study (y-direction) and do not reflect any of the systematic uncertainties.}
\label{fig:spec_comp}
\end{figure*}

In this section we compare our spectroscopic line ratio measurements to those determined from narrow-band photometry given in Table 4 of MR95. \citet{Dutil2001} point out some of the hazards of using imaging spectrophotometry for studying chemical abundances across extended objects such as galaxies. They demonstrate that the main limit to their method is the subtraction of the underlying continuum. We compare our high S/N spectroscopic measurements with robust measurements of the underlying continuum with the spectrophotometric measurements of MR95 in order to asses any systematic bias that may result from photometric determinations. 

We match HII regions in common with MR95 by eye, comparing our image shown in Figure \ref{fig:img} to a similar figure (Figure 3) in MR95. We note that for 2 of the 17 matched HII regions there is ambiguity owing to the high density of labeled points in MR95. This ambiguity is inconsequential for one of the regions because the line ratios for the two possible matches are consistent to within the errors. For the second case, we select the region that is most consistent with the observed relationship between the spectroscopic and spectrophotometric measurements inferred from the other 16 HII regions. The selection has very little effect on the analysis and no effect on the conclusions.

Figure \ref{fig:spec_comp} shows the comparison for the $[OIII]/H\beta$ (left), $[NII]/[OIII]$ (center) and $[NII]/H\alpha$ (right) line ratios. The errors on the MR95 data are taken from their measurements. The 17 HII regions span the full metallicity range covered by the MR95 data. The solid line in each of the plots is the one-to-one agreement and the dashed line is a least-square fit to the relationship. The dispersion around the fitted relation is 0.12, 0.18 and 0.10 for $[OIII]/H\beta$, $[NII]/[OIII]$ and $[NII]/H\alpha$, respectively. The $[NII]/[OIII]$ line ratio has the largest dispersion which we attribute to the additional scatter resulting from uncertainty in the reddening correction due to those lines being $\rm \sim\!1600~\AA$ apart. 

\citet{Dutil2001} show that the spectrophotometric method is limited by the continuum subtraction. The systematic deviation of the MR95 data from our data, highlighted by the linear fits in Figure \ref{fig:spec_comp}, are consistent with a systematic error in the subtraction of the continuum. We expect a systematic error in the continuum subtraction to most strongly effect the weaker lines. For the ratios shown in Figure \ref{fig:spec_comp}, the weaker line is $H\beta$ for the $[OIII]/H\beta$ line ratio (left panel) and $[NII]$ for the $[NII]/[OIII]$ and $[NII]/H\alpha$ line ratio (center and right panel, respectively). The deviation observed is consistent with an overestimate of the continuum leading to an underestimate of the line strengths in the MR95 data.

We derive a linear correction for each of the line ratios of the photometric data of MR95 from the 17 measurements displayed in Figure \ref{fig:spec_comp} by minimizing the square of the residuals, weighted by the statistical errors, between the two determinations and apply this correction to all 77 line ratio measurements made by MR95. Hereafter, when referring to the data of MR95, we mean to refer to these corrected data.

The linear fits shown in each of the three panels of Figure \ref{fig:spec_comp} were performed using the routine \emph{fitexy.pro} in IDL. The routine does a linear least-square approximation using the errors in both the x and y data. The fits are given by:
\begin{equation}
S_{O3H\beta} = (-0.08 \pm 0.02) + (1.32 \pm 0.08) P_{O3H\beta}, 
\label{eq1}
\end{equation}
\begin{equation}
S_{N2O3} = (-0.01 \pm 0.02) + (1.24 \pm 0.04) P_{N2O3}, 
\end{equation}
and 
\begin{equation}
S_{N2} = (0.12 \pm 0.02) + (1.21 \pm 0.07) P_{N2}.
\label{eq3}
\end{equation}
Here S and P denote the spectroscopic and photometric data, respectively. The subscripts in each of the equations denote the line ratio in consideration.  The errors in the linear fits in equations \ref{eq1} - \ref{eq3} reflect measurement uncertainties but do not account for systematic uncertainties.

We note that, to within the scatter, the line ratios appear to have only linear systematic differences. Furthermore, the abundances determined from the \citet{Pettini2004} diagnostics are nearly linear in metallicity with respect to the line ratio used. Therefore, it is extremely unlikely that an abundance gradient break could spuriously result from the use of the photometric as opposed spectroscopic line ratios. 

We conclude that given the greater systematic uncertainties in the spectrophotometric method of line ratio measurement, our spectroscopic data provide a more robust measure of the line ratios used in ionized gas diagnostics as compared with the monochromatic imaging data of MR95. However, the strong correlation and relatively small dispersion of MR95 with our data give us confidence in the corrected data of MR95 and its use in estimating abundances.

\section{Abundance Gradient Break}

\begin{figure*}
\includegraphics[width=7in]{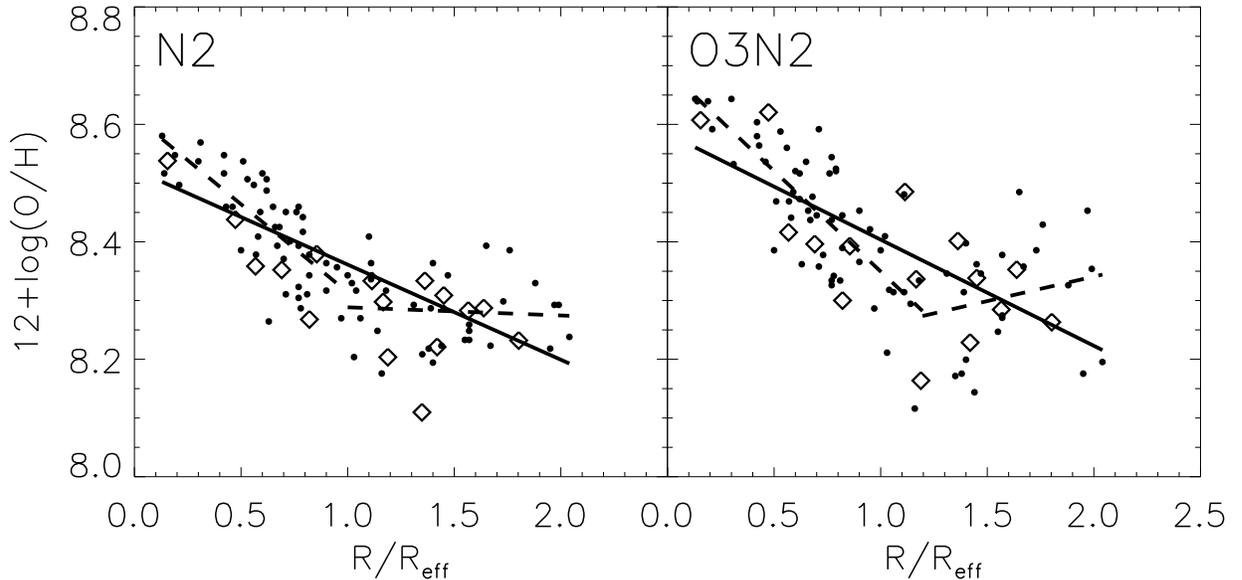}
\caption{The abundance gradient for NGC 3359 derived from $[NII]/H\alpha$ (left) and $[OIII]/[NII]$ (right). The black data points are the metallicities determined from the $N2$ and $O3N2$ diagnostics of \citet{Pettini2004} using the line ratios of MR95. The diamonds are the metallicities determined for 19 regions from our spectroscopic measurements of the line ratios using the same metallicity diagnostics. The solid line in each panel is a linear model fit to the MR95 data and the dashed lines in each panel is a model with a break fitted to the MR95 data.}
\label{fig:ftest}
\end{figure*}
%\begin{deluxetable}{lcccccc}
%\tablewidth{\columnwidth}
%\tablecaption{F-test Statistic}
%\tablehead{{Diagnostic} & \colhead{$R_1$} &\colhead{$R_2$} & \colhead{$P_1$}& \colhead{$P_2$}& \colhead{$N$}& \colhead{$F$}  } 
%% All data must appear between the \startdata and \enddata commands
%\startdata
%N2       & 0.247 & 0.177 & 2 & 5 & 77 & 9.416 \\
%O3N2 & 0.592 & 0.426 & 2 & 5 & 76 & 9.206 \\
%\enddata
%\label{tab:f}

%\tablecomments{The first column gives the diagnostic used to derive metallicity. R and P are the sum of the residuals squared and the number of parameters in each model. The subscripts denote model 1 and model 2, respectively, where model 1 is the two parameter linear model and model 2 is a five parameter two-line model plus one parameter for the break. The final column is the F-statistic as defined by equation \ref{eq:f}.}
%\end{deluxetable}

The oxygen abundance gradient in NGC~3359, first investigated by MR95, offers one of the clearest examples of a break in the radial gradient found for barred spiral galaxies. In this section, we reanalyze the data of MR95, but employ a more robust statistical approach for establishing the break. The data of MR95 are consistent with our observations and we have recalculated the metallicities using the empirically calibrated diagnostics of \citet{Pettini2004}. The $N2$ ratio does not suffer from the degeneracy found in the oxygen line ratios due to its monotonic behavior with respect to metallicity. Additionally, due to the very close spacing of the the two lines involved, the ratio does not require any corrections for reddening. We also calculate metallicities based on the $O3N2$ line ratio, though this diagnostic yields a considerably larger scatter in $O/H$.

Figure \ref{fig:ftest} shows the abundance gradient derived from the $N2$ (left) and $O3N2$ (right) diagnostics. The black data points are the metallicities determined from the data of MR95. The diamonds are the metallicities for 19 HII regions derived from our spectroscopic observations. The solid line in each plot is a least square fit to the data from MR95 while the dashed lines are a linear model with a break fit to the same data. The model with a break  has five parameters, four of which define slopes and intercepts of two lines, while the fifth parameter defines the location of the break in units of the effective radius. MR95 take the break to occur at an effective radius of $R_{\mbox{\em\scriptsize eff}} = 1$.

In order to establish the significance of a model with a break in the gradient we perform a likelihood ratio F-test. This test is appropriate when the distribution of the statistic is a F-distribution under the null hypothesis. This test is most commonly used when comparing nested statistical models such that the model with fewer parameters, model 1, can be reproduced by fixing parameters in the model with greater number of parameters, model 2. In our case, model 1 (linear model) can be reproduced by setting the break to be at $R_{\mbox{\em\scriptsize eff}}>R_{max}$, where $R_{max}$ is the maximum effective radius of the data, making the second line inconsequential when determining the sum of the residuals squared in model 2 (model with a break). Model 2 will always fit as well, if not better than model 1. The purpose of performing the F-test is to judge whether the decrease in the sum of the residuals squared is significant given the increase in the number of parameters.

One can calculate the F-statistic for the null hypothesis that model 2 does not give a significantly better fit than model 1. The F-statistic is given by
\begin{equation}
F = \frac{\frac{\chi_1^2 - \chi_2^2}{P_2 - P_1}}{\frac{\chi_2^2}{N-P_2}}.
\label{eq:f}
\end{equation}
Here, 
\begin{equation}
\chi^2 = \sum \frac{(x_i - m_i)^2}{\sigma_i^2},
\end{equation}
where $x_i - m_i$ are the residuals between the model and the observations and $\sigma_i^2$ are the errors in the observations. $P$ is the number of parameters and the subscripts denote the model.  N is the number of data points. Our derived F values are 8.5 and 13.6 for the $N2$ and $O3N2$ diagnostics, respectively. Under the null hypothesis, F has a F-distribution with $(P_2 - P_1, N - P_2)$ degrees of freedom. We can derive the probability at which we can reject the null hypothesis by comparing our calculated F value with a critical value at some false-rejection probability. For F distributed with (3, 72) degrees of freedom, the critical value is 8.1 for a false-rejection probability of 0.0001. The calculated F values suggest that the probability that we would falsely reject the null hypothesis is less than $0.01\%$. To put it in terms of a confidence, we can say with over $99.99\%$ confidence that the null hypothesis is rejected, meaning that model 2 is a significantly better than model 1 at fitting the data. A F-test performed on our spectroscopic data alone gives a confidence of $75\%$ and $65\%$ for the $N2$ and $O3N2$ diagnostics, respectively. 

From visual inspection of the left panel of Figure \ref{fig:ftest} it can be seen that the gradient does not appear to be continuous. The greater scatter in the right panel makes this less obvious. Our spectroscopic measurements, displayed by the diamonds in Figure \ref{fig:ftest}, have similar scatter to the measurements of MR95 and help to corroborate the presence of break, especially when using the $N2$ diagnostic. The statistical analysis of the data of MR95 suggests that for both metallicity diagnostics a model with a break provides a significantly better characterization of the data. The same statistical tests performed on our spectroscopic data suggest that a model with a break provides only a marginally better fit to the data. The significantly lower confidence results from the substantially fewer degrees of freedom in our fit owing to the greatly reduced sample size.

The location of the break, in units of $R_{\mbox{\em\scriptsize eff}}$, is 1.0 and 1.2 for the $N2$ and $O3N2$ diagnostics, respectively. The location of the break is in rough agreement between the two diagnostics and the value derived from the $N2$ diagnostic is exactly consistent with MR95. For both metallicity diagnostics, the slope of the outer gradient, within the errors, is consistent with zero. We do not expect the effective radius to play any significant role in determining the gas dynamics driving the observed break in the abundance gradient. From numerical simulations of NGC 3359, \citet{Rozas2008} find that the corotation radius is  $95''$. The effective radius for this galaxy is $87''$ and we attribute the break location of $R_{\mbox{\em\scriptsize eff}} = 1$ to the near coincidental location of the effective and corotation radii. 

\section{Ionization Parameter and the R23 Diagnostic}

\begin{figure}
\includegraphics[width=\columnwidth]{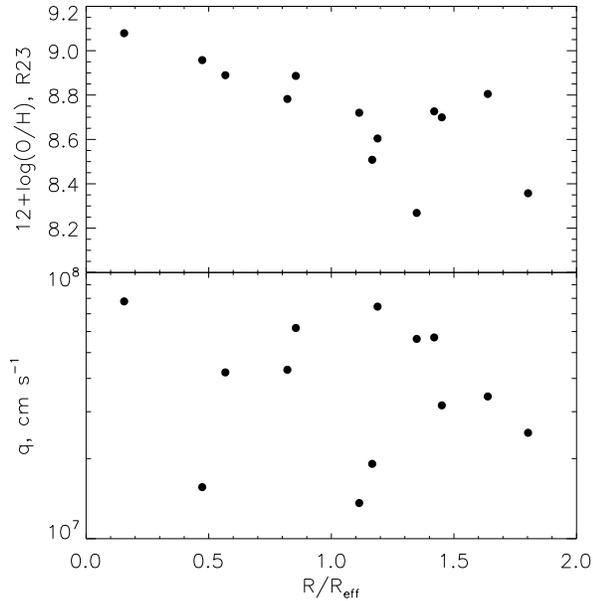}
\caption{The abundance (top panel) and ionization parameter, q, (bottom panel) derived from the $R23$ and $O32$ line ratios, respectively, using the scheme of \citet{Kobulnicky2004}. The ionization parameter is in units of cm s$^{-1}$. The upper/lower branch degeneracy is broken using the $[NII]/[OII]$ line ratio \citep{Kewley2008}.}
\label{fig:r23}
\end{figure}

We determine the oxygen abundance and ionization parameter from the $R23$ and $O32$ line ratios, respectively, using the calibration of \citet{Kobulnicky2004}. The line ratios are defined such that
\begin{equation}
R23 = log\{([OII] + [OIII])/H{\beta}\}
\label{eq:r23}
\end{equation}
and 
\begin{equation}
O32 = log\{[OIII]/[OII]\}.
\label{eq:o32}
\end{equation}
Where $[OIII]$ is the sum of the line flux of $[OIII]\lambda4959$ and $[OIII]\lambda5007$ and $[OII]$ is the flux of the doublet centered at $3727\AA$. This diagnostic builds on the photoionization grids and stellar evolution models of \citet{Kewley2002}.

The $R23$ method for determining metallicities is known to depend on the ionization parameter. \citet{Pilyugin2003} has argued that artificial breaks in the abundance gradient may result from radial variations in the ionization parameter. The ionization parameter, which we denote by q, characterizes the ionization state of the gas within the HII region and quantitatively represents the number of ionizing photons per second per unit area divided by the hydrogen number density. The ionization parameter has the units of velocity and physically represents the maximum velocity of the ionization front through the nebula. The metallicity and ionization parameter are interdependent. We use equations A4 and A6 given in the appendix of \citet{Kewley2008} to determine the ionization parameter and metallicity respectively. In order to obtain a consistent measurement of the metallicity and ionization parameter an iterative scheme is used, the details of which are provided in the appendix of \citet{Kewley2008}.

Metallicity is not a singly-valued function of $R23$, but rather takes on two values for a given ratio. The peak of the $R23$ ratio occurs at $12 + log(O/H)\sim8.4$. This degeneracy is due to the fact that on the lower branch $R23$ increases with metallicity because the intensity of the collisionally excited [OII] and [OIII] lines increase with the abundance. On the upper branch, nebular cooling, which results from a collisional excitation followed by photon emission, effectively cools the nebula decreasing the electron temperature leading to a decrease in the rate of collisional excitation of the [OII] and [OIII] lines. In order to break this degeneracy we use the $[NII]/[OII]$ line ratio. We place objects on the lower branch of the $R23$ diagnostic if  $log([NII]/[OII]) < -1.2$ \citep[see appendix of][]{Kewley2008}.

Figure \ref{fig:r23} shows the oxygen abundance and ionization parameter determined from the iterative scheme of \citet{Kobulnicky2004} using the $R23$ and $O32$ line ratios, respectively, plotted against effective radius. We were able to determine the metallicities in only 13 of 19 HII regions due to the non-detection of the [OII] doublet, 3 of which are found to be on the lower branch. MR95 do not observe the $[OII]$ line so we are unable to perform this diagnostic on their data.

The abundance gradient shown in the top panel of Figure \ref{fig:r23} is suggestive of a break around $R/R_{eff} \sim 1$. More importantly, \emph{there is no observed trend in the ionization parameter with radial distance}. This provides strong evidence against the assertion that the break in the abundance gradient may be attributable to the radial variations in the ionization parameter.

\ \\

\section{Discussion and Conclusions}

We build on the work of \citet{Martin1995} in order to validate the radial oxygen abundance gradient break they report. We find that the photometric determination of line ratios agrees reasonably well with the spectroscopic line measurements. We make a small correction for systematic differences between the two methods of measurement which we attribute to uncertainties in continuum subtraction and reddening corrections. Both the corrected data of MR95 and our spectroscopic measurements of 19 HII regions show evidence of a break in the gradient. We statistically compare a model with a single line to a model with a break. We find that the model with a break provides a significantly better characterization of the data. We determine the ionization parameter from the $O32$ line ratio and find that it appears to be uncorrelated to the radial distance.

\citet{Pilyugin2003} examines the observed bends in the slopes of the radial oxygen abundance gradients in spiral galaxies. He argues that the upper/lower branch degeneracy in the metallicity derived from $R23$ along with gradients in the ionization parameter may result in a false break or bend in the radial oxygen abundance gradient. The main focus of this work is to address these issues.

In order to address the $R23$ degeneracy, we use an abundance indicator that is calibrated for these type of data, namely the spectroscopically determined $N2$ line ratio measurements of individual HII regions in the metallicity range $8.6\gtrsim12+log(O/H)\gtrsim8.2$ \citep{Pettini2004}. Because the $N2$ ratio is a monotonic function of metallicity, it avoids the degeneracy issues that arise when using $R23$. Additionally, for 13 of our 19 HII regions we were able to break the degeneracy in the $R23$ diagnostic using the $[NII]/[OII]$ line ratio. The abundance gradient determined from the $R23$ diagnostic appears to have a break consistent with the determination from the $N2$ and $O3N2$ diagnostics, though it cannot be established from these data alone due to the small number of regions observed. 

The second issue raised by \citet{Pilyugin2003} is that a radial oxygen abundance gradient break can result from a radial variation in the ionization parameter. He shows that for NGC 2403 a false bend results from using the $R23$ method for metallicity determination. We compile data from the literature for NGC 2403 and determine the metallicity using the $N2$ diagnostic \citep{McCall1985, Fierro1986, Garnett1997, vanZee1998}. We find no evidence of a break. This suggests that the $N2$ diagnostic is less susceptible to such confusion. Moreover, we examine the ionization parameter in NGC 3359 and find no apparent correlation with radial distance. We conclude that the break in the radial abundance gradient we observe in NGC 3359 is not due to a radial variations in the ionization parameter. In this way, we are able to statistically establish the abundance gradient break in NGC 3359 while avoiding the problems associated with the points of concern raised by \citet{Pilyugin2003}.

The radially differential star formation rate is taken to be the cause of abundance gradients in spiral galaxies. However, it has been observed that barred galaxies generally have shallower gradients due to large scale gas mixing associated with the bar formation. The presence of breaks in the abundance gradients is consistent with numerical simulations of barred galaxies with a young, still-forming bar \citep{Friedli1994, Friedli1995}. It is thought that because the bar is still forming and there is ongoing star formation, turbulent gas mixing has not had the time to smooth out the radial abundance gradient inside the corotation radius, while outside the corotation radius, large-scale gas motions have homogenized the abundance. 

The formation and presence of a bar has been associated with morphological transformations. \citet{Dutil1999} argue that bars may be able to transform late-type galaxies into early-type ones. They find that the central abundance gradient of galaxies when plotted against morphological type is suggestive of two distinct sequences, one formed by late-type barred and early-type (barred and unbarred) galaxies and another comprised of normal late-type galaxies. One interpretation of this sequence is that the bar flattens the abundance gradient and dissolves with time in early-type galaxies. In simulations, the dissolution of the bar has been linked to the growth of the bulge and could be responsible for morphological transformations \citep{Combes1990, Norman1996, Pfenniger1998}. However, the observational evidence for bar dissolution is limited \citep{Block2002, Das2003}. 

%In this context, a break in the abundance gradient, suggesting a young bar in the process of formation, may be taken as evidence for a morphological transformation in progress.

From recent infrared studies, the fraction of local spiral galaxies having bars is well established and found to be around $\sim60\%$ \citep{Eskridge2000, Marinova2007, Menendez-Delmestre2007}. \citet{Sheth2008} find that the bar fraction for spiral galaxies is greater for higher stellar mass systems and that this relation evolves from $0.2<z<0.8$, with the greatest evolution in the fraction occurring for the lowest mass systems. This suggests that galaxy downsizing is not only a process of star formation regulation in massive systems but also one of dynamical evolution. 

The presence of bars in local spiral galaxies is prevalent while the epoch of bar formation is short. Therefore, observations of galaxies where such abundance gradient breaks are expected to occur are rare. Bar formation may be a process leading to morphological transformations and has been shown to exhibit downsizing.  Abundance gradient breaks have been taken as evidence of the young bars still in the process of formation. Observing and understanding this transient epoch of formation may prove to be a very important aspect of understanding the processes governing the evolution of galaxies.

\acknowledgments
FB and HJZ gratefully acknowledge support by NSF grant AST-0707911.  We acknowledge the cultural significance Mauna Kea has for the Hawaiian community and with all due respect say mahalo for its use in this work.

\bibliographystyle{apj}

\bibliography{gradient.bib}

 \end{document}